\documentclass[a4paper,fleqn]{cas-sc}

\usepackage[numbers]{natbib}

\usepackage{soul}

\usepackage{url}

\usepackage[ruled,vlined]{algorithm2e}
\usepackage{algorithmic}
\SetKwInput{KwInput}{Input}
\SetKwInput{KwOutput}{Output}

\usepackage{ragged2e}

\def\tsc#1{\csdef{#1}{\textsc{\lowercase{#1}}\xspace}}
\tsc{WGM}
\tsc{BEC}

\begin{document}
\let\WriteBookmarks\relax
\def\floatpagepagefraction{1}
\def\textpagefraction{.001}
\shorttitle{GOES GLM Bolide Detection Pipeline}
\shortauthors{J.C. Smith et~al.}

\title [mode = title]{An Automated Bolide Detection Pipeline for GOES GLM}

\author[1,2]{Jeffrey C. Smith}[type=author,
                        orcid=0000-0002-6148-7903]
\cormark[1]

\credit{Machine Learning Training, Pipeline Architecture, Data Analysis}

\address[1]{The SETI Institute, 189 Bernardo Ave, Suite 200, Mountain View, CA 94043, USA}
\address[2]{NASA Ames Research Center, Moffett Field, CA 94035, USA}

\author[1,2]{Robert L. Morris}

\credit{Pipeline Arhitecture}

\author[2, 3]{Clemens Rumpf}[orcid=0000-0003-2530-4046]
\address[3]{STC, NASA Research Park, Moffett Field, CA 94035, USA}

\credit{Prototype Algorithm Development}

\author[4]{Randolph Longenbaugh}

\credit{Bolide Analysis and Vetting}

\address[4]{Sandia National Laboratories, Albuquerque, NM 87185, USA}

\author[2]{Nina McCurdy}

\credit{Visualization}

\author[2]{Christopher Henze}

\credit{Data Management}

\author[2]{Jessie Dotson}[orcid=0000-0003-4206-5649]

\credit{Project Management}


\cortext[cor1]{Principal corresponding author}

\begin{abstract}
The Geostationary Lightning Mapper (GLM) instrument onboard the GOES 16 and 17 satellites has been shown to be capable
of detecting bolides (bright meteors) in Earth's atmosphere. Due to its large, continuous field of view and immediate
public data availability, GLM provides a unique opportunity to detect a large variety of bolides, including those in
the 0.1 to 3 m diameter range and complements current ground-based bolide detection systems, which are typically
sensitive to smaller events. We present a machine learning-based bolide
detection and light curve generation pipeline being developed at NASA Ames Research Center as part of NASA’s
Asteroid Threat Assessment Project (ATAP).  The ultimate goal is to generate a large catalog of calibrated bolide lightcurves
to provide an unprecedented data set which will be used to inform meteor entry models on how incoming bodies
interact with the Earth's atmosphere and to infer the pre-entry properties of the impacting bodies. The data set will
also be useful for other asteroidal studies. This paper reports on the progress of the first part of this ultimate goal, 
namely, the automated bolide detection pipeline. 

Development of the training set, ML model training and iterative improvements in detection performance are presented.
The pipeline runs in an automated fashion and bolide lightcurves along with other measured properties are promptly
published on a NASA hosted publicly accessible website, https://neo-bolide.ndc.nasa.gov.

\end{abstract}


\begin{highlights}
\item GOES Geostationary Lightning Mappers can be used to detect bolides. 
\item A Random Forest detection algorithm has been deployed on the NASA Advanced Supercomputing facility Pleiades supercomputer.
\item Bolide light curves are promptly published on a NASA hosted publicly accessible website, https://neo-bolide.ndc.nasa.gov.
\end{highlights}


\maketitle


\section{Introduction}
\label{s:intro}

There currently exists no large, uniformly processed set of bolide light curves. The uniformly processed bolide light
curves that do exist are of limited atmospheric collecting areas and total only hundreds of bolides
\citep{1996M&PS...31..185H, 1997M&PSA..32..157C}. We aim to increase the set of uniformly processed bolide light curves
by over an order of magnitude.  NASA's Asteroid Threat Assessment Project (ATAP), a NASA Ames Research Center activity
in support of NASA’s Planetary Defense Coordination Office (PDCO), is generating a database of bolide light curves and
is currently developing a machine learning based detector using GOES GLM satellite data. A prototype has been running
for over a year now with well over 1700 bolides posted at: https://neo-bolide.ndc.nasa.gov. Our goal is the detection,
characterization and generation of a large dataset of calibrated entry light curves of bolides in order to calibrate our
entry models to asses the risk of larger meteor impacts, but the publicly available data set is useful for numerous
other science studies, including asteroid population studies to better understand the evolution of our Solar System.

ATAP desires such a data set for use with tuning entry modeling software, which currently uses a very small number of
well studied events, such as the Chelyabinsk meteor \citep{brownChelyabinsk2013, Popova1069, borovickaChelyabinsk2013}.
The bolide detection projects that do exist mainly use ground-based camera systems which have limited sight and easily
saturate and thus do not produce accurate disintegration light curves. The data set we are collecting does not
experience these limitations. Also, with the stereo detection nature of GOES 16 and 17 (i.e.  the region where GOES 16
and 17 view the same region of the globe), it is also possible to reconstruct the trajectory of the bolides, which
greatly aids in the bolide reconstruction modelling. However, even in the non-stereo regions, the high fidelity of the
brighter bolide event light curves means there is great potential to still determine much about the events. The
trajectory information will also be useful for both understanding the object's origin (i.e. Planetary Science) and final
destination (for possible recovery). 

There do exist numerous bolide detection programs. Some bolides are found using space-based United States Government
sensors (USG) \citep{TagliaferriInBookBolides}.  They have a large collection area and are sensitive to large events,
but due to national security interests, the data sets released are limited and only after a significant time delay.  The
published detections from these sources include the geo-located position and time of peak brightness, as well as the
cumulative energy release, but no light curves. Other bolide detection sources include
ground-based meteoritic observatories, such as the European Fireball Network \citep{2012epsc.conf..441F}, Desert
Fireball Network \citep{desertFireballNetwork, 2017ExA....43..237H}, NASA Meteorite Tracking and Recovery Network
and NASA All Sky Fireball Network \citep{NASAMetTrakRecNetwork, NASAAllSkyFIreNetwork, 2012pimo.conf....9C}, Sky Sentinel
\citep{skySentinal}, Cameras for Allsky Meteor Surveillance (CAMS) \citep{CAMS, 2011Icar..216...40J}, Spanish fireball network
\citep{8136678} and Fireball Recovery and Interplanetary Observation Network (FRIPON) \citep{FRIPON, 2020A&A...644A..53C}.  
Many are members
of the extensive Global Fireball Observatory \citep{globalFireballObservatory, 2020P&SS..19105036D}.  
These latter sources have relatively small atmospheric collecting areas
but when used in aggregate can provide a large collection area.  Nonetheless, they still do not compete with the
hemispherical view of the Earth available to the GOES GLM instruments. They are sensitive to smaller (dimmer) events than
GLM but saturate if the bolide becomes too bright. Furthermore, due to the distributed nature of all the ground-based
sources and that they are maintained by different scientific groups, the data products are not necessarily uniform or
have similar calibrations.  A fully separate method to detect bolides involves using infrasonic detection 
\citep{edwardsInfrasonic,silberInfrasonic}, such as by the Comprehensive Nuclear Test Ban Treaty Organization, CTBTO
\citep{dahlmanCTBTO}. GOES GLM data complements all these others sources. GLM provides a fast ultra-wide field of view
(encompassing two continents and oceans over the western hemisphere, as seen in Figure \ref{f:websiteBolidesOnGlobe}),
sensitivity to moderately sized events (0.1 to 3 m diameter \citep{rumpf-glm}), and uniform data products, providing an
unprecedented data set for bolide studies.

The pipeline work has been carried out in a series of stages. Stage 1 was to develop a prototype detector using classic
filter design \citep{rumpf-glm} and demonstrate the feasibility of detecting bolides in GLM. Stage 2 was to wrap the
prototype algorithms in a robust pipeline architecture for automatic and scheduled execution on the NASA Advanced
Supercomputing (NAS) facility Pleiades Supercomputer \citep{NAS} at NASA Ames Research Center. Stage 3 was to collect
detections from the prototype pipeline and then perform human vetting in order to generate a labeled data set. Stage 4
was to use the training data set to train a machine learning classifier to improve the performance of the detector. As
is typically the case with supervised machine learning algorithms, the reliability of the detections in the automated
pipeline is reliant upon the reliability of the human vetting which generated the training data set. 

\section{GOES GLM Data}

The GOES Geostationary Lightning Mappers \citep{2013AtmRe.125...34G} positioned over the western hemisphere in
geosynchronous orbit detect transient light events at 500 frames per second.  The GLM instrument is a staring CCD imager
(1372 $\times$ 1300 pixels) which measures emissions in a narrow 1.1 nm pass-band centered at 777.4 nm, a principal
wavelength for the neutral atomic oxygen emission (OI bound-bound) triplet line of the lightning spectrum. The pass-band
is best for distinguishing lightning against a bright daytime cloud background. It is quite narrow,
nonetheless, within this bandpass GLM will detect continuum emissions \citep{doi:10.1111/maps.13137} and bolides emit
continuum radiation as well as O triplet lines. This means despite GLM's singular goal of detecting lightning, papers
have already demonstrated GLM's ability to detect bolide events \citep{doi:10.1111/maps.13137, rumpf-glm}. 

Although bolide emission is not identical to lightning, it is approximately correct to consider a bolide spectrum to be
the sum of a continuum component (that is approximately flat within the 1.1nm pass-band) and the oxygen triplet.  The
``shock layer'' (the region between the surface of the meteoroid and the bow shock upstream of it) is normally a very
bright continuum source, and much brighter than the oxygen triplet, however the GLM instrument observes the data from
above and therefore the bulk mass of the bolide blocks a direct view of the bow shock. In contrast, the Oxygen triplet
emission in the tail has a direct line of sight with the GLM instrument and is therefore a potentially significant
contribution to the overall flux emitted by the event as observed by GLM. The relative weights of these two first-order
components (continuum and oxygen triplet) are determined primarily by object diameter, altitude, velocity, and view
angle. Second order effects include the treatment of ablation physics and requires detailed knowledge of the asteroid
composition \citep{JOHNSTON2021114037}. We suffice it to say, exact bolide emission within the GLM bandpass is an area
of active research.

Retrieving the full frame images for GLM every 2 ms would far exceed the 7.7 mbps telemetry downlink rate. Onboard
processing is therefore necessary to identify and bring down just the lightning events. Downlinked GLM data is organized
into \emph{events}, \emph{groups} and \emph{flashes}.  An \emph{event} is defined as the occurrence of a single pixel
exceeding the background threshold during a single frame.  A \emph{group} is one or more simultaneous events that
register in adjacent (neighboring or diagonal) pixels in the focal plane array. A single lightning discharge will
typically illuminate multiple pixels.  A \emph{group} can be considered to be a single lightning pulse, but it is
possible that multiple pulses occur within the 2 ms integration window.  A \emph{flash} consists of one or more groups
within a specified time and distance defined as a set of groups sequentially separated in time by 330 ms or less and in
space by no more than 16.5 km (nominally two pixels) in a weighted Euclidean distance format. Note that a train of
groups with total distance greater than 2 pixels can contribute to a flash provided they are part of a contiguous train.
The lightning flash detection efficiency is 70\% with 5\% false alarms. Bolides are an exceedingly small fraction of the
false alarms.

The raw GLM Level 0 data products are then processed through the ground-segment pipeline to collect, filter, geo-locate
and time tag the raw data into Level 1B lightning events and then through the GLM Lightning Cluster-Filter Algorithm
into Level 2 data products \citep{GLM-ATBD}.  The vast majority of events in the Level 2 data products is lightning
activity, as would be expected, but other events can also be observed. For our purposes we label GLM events into three
categories: 1) lightning, 2) glints/instrumental and 3) bolides. Naturally, we are only interested in the latter, the
former two are considered background.  We currently use the Level 2 group data for bolide detection.

\begin{figure}
	\centering
		\includegraphics[scale=.25]{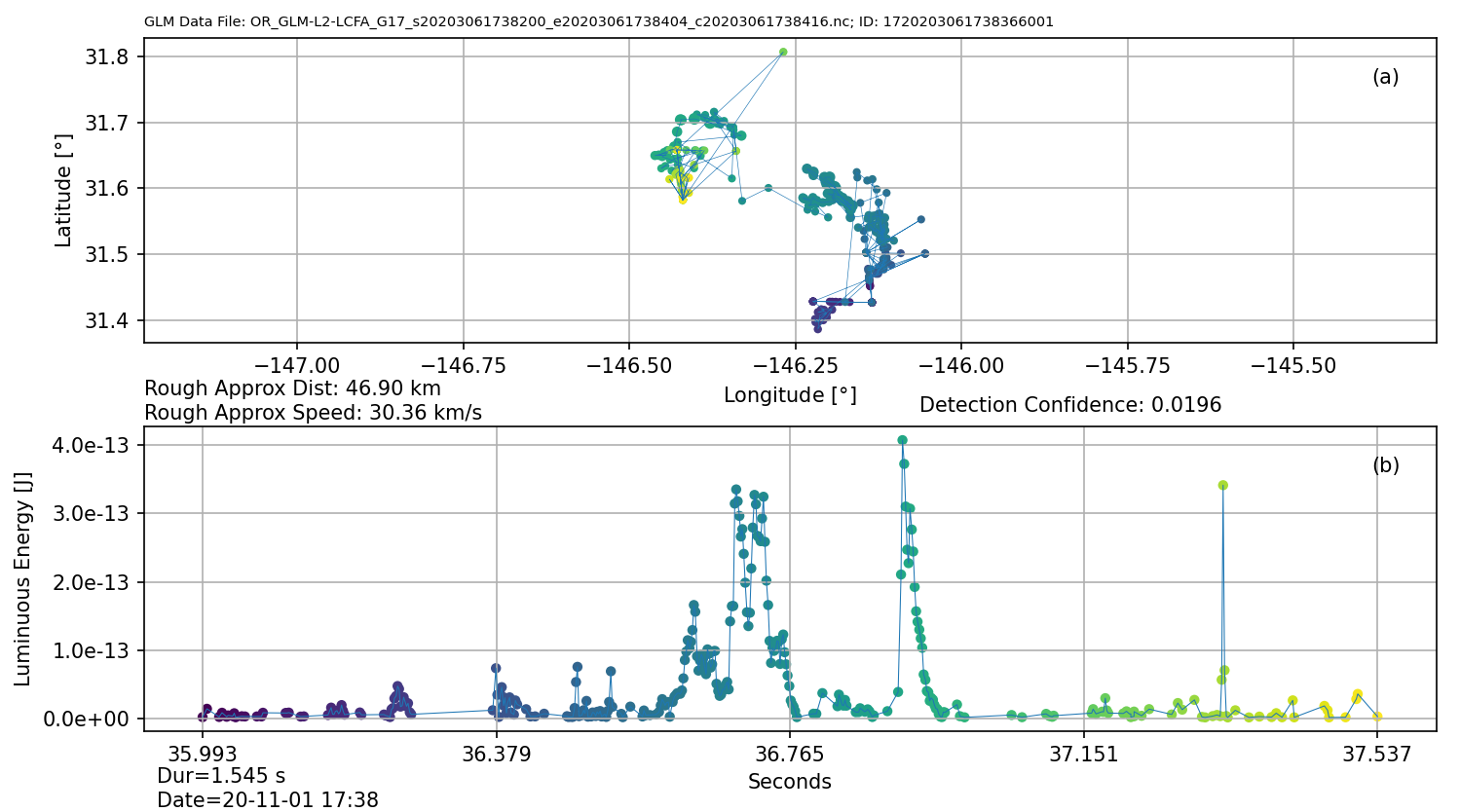}
        \caption{GLM group data example of typical lightning activity. The top plot (a) gives the GLM reported ground track in
        latitude and longitude. The bottom plot (b) gives the GLM reported luminous energy in Joules. Time runs from
        blue to yellow.}
	\label{f:lightningExample}
\end{figure}

\begin{figure}
	\centering
		\includegraphics[scale=.25]{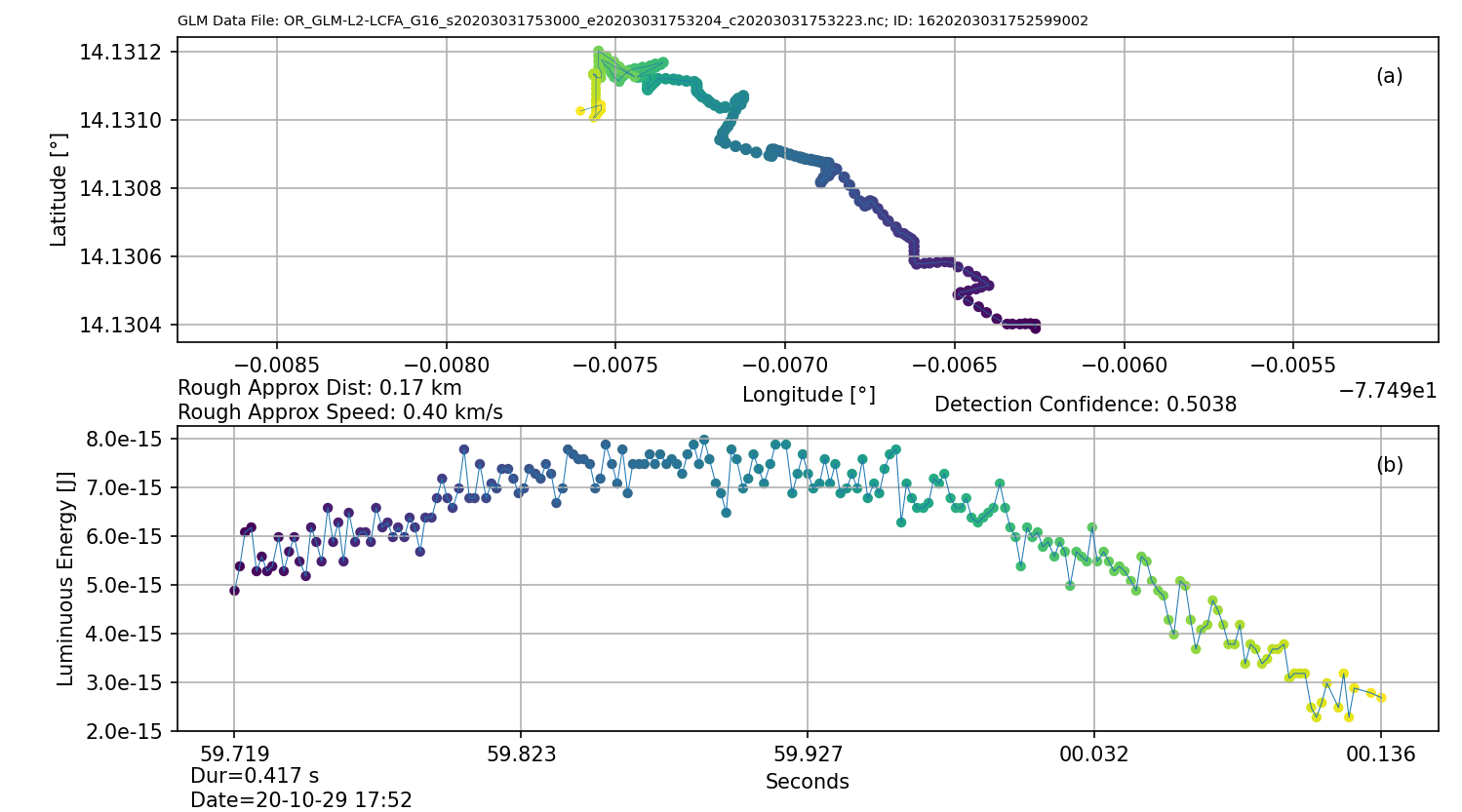}
        \caption{GLM group data example of typical glint activity.}
	\label{f:glintExample}
\end{figure}

\begin{figure}
	\centering
		\includegraphics[scale=.25]{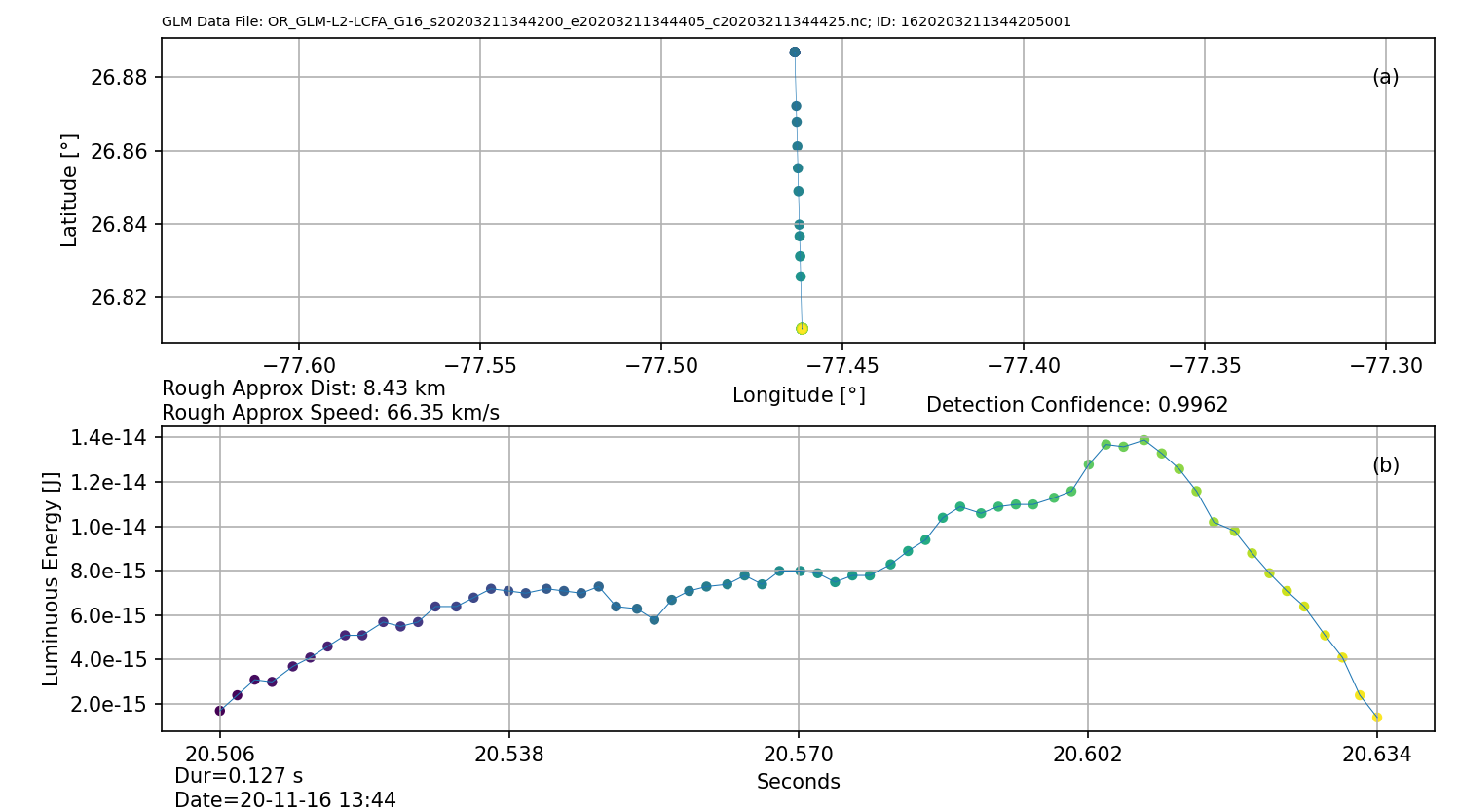}
        \caption{GLM group data example of a legitimate bolide.}
	\label{f:bolideExample}
\end{figure}

Figures \ref{f:lightningExample}, \ref{f:glintExample} and \ref{f:bolideExample} give examples of GLM group data for
each of the three categories of events. Figure \ref{f:lightningExample} is an example of typical lightning activity. The
light curve does not have a smooth trend, instead it is very stochastic with abrupt flashes. The ground trace is also
very scattered with no clear direction of motion for the event. Figure \ref{f:glintExample}  is a representative example
of a glint event, which are reflections of the sun off the Earth's surface. They tend to extend long in time and will
sometimes exhibit a long smooth symmetrical trend in the light curve but the ground track will meander.  The meandering
is commonly indicative of reflection off a body of water, where the brightest point may be due to surface wave
undulations and regional reflectivity variations.  Figure \ref{f:bolideExample} is a legitimate bolide event that
occurred on November 16th, 2020 during the Leonid meteor shower. Here we see the distinctive signature of a bolide. As
the meteor enters and begins to interact with the atmosphere it will heat up and become luminous.  The brightness will
continue to increase, generally gradually, until peak brightness occurs when the bolide begins to fragment. The light
will then usually fade rapidly relative to the build-up and the total time of the event is a small fraction of a second,
in this case at 0.127 s, but sometimes can last longer.  Also sometimes, multiple fragmentation events will occur. The
ground track is also very linear, as would be expected for a fast meteoroid entering the atmosphere.  Based on the
ground track, this object travelled at a velocity of $66$ km/s, consistent with a Leonid event \citep{AMS2020Meteors},
however there are large uncertainties in the measured velocity due to perspective.
The data plotted in the figures is that reported by the GLM Level 2 data products. No uncertainties are reported in
these data products, and therefore no error bars are included in the figures.

The design of the GLM instruments and both their onboard and ground data processing algorithms, which are tailored to
monitor lightning, pose a number of significant challenges to recovering calibrated bolide light curves. The GLM group
light curves thus produced and illustrated in Figures \ref{f:lightningExample}, \ref{f:glintExample} and
\ref{f:bolideExample} are therefore not properly calibrated for bolide events. In particular, GLM uses adaptive event
thresholding and background estimation algorithms which can cause a significant fraction of the measured bolide flux to
be discarded. Recovery of flux in calibrated units is further complicated by the fact that the object's altitude,
velocity, size, and view angle all affect the spectral irradiance at the GLM detector. The reported GLM group time is
corrected for light time but to typical lightning altitudes of approximately 10km \citep{10.1117/12.2242141},
whereas bolides are typically around 80km \citep{doi:10.1111/maps.13137}. A parallel effort within ATAP aims to recover calibrated light
curves from the Level 0 data products, including uncertainties.  We will leverage knowledge about the instrument's
optics, the onboard data processing, and bolide emissions, and the work will be presented in a separate paper.

\section{Initial Pipeline Architecture, Classic Filter Design and Manually Vetted Data}

NASA's Asteroid Threat Assessment Project (ATAP) has been developing techniques to detect bolides in GLM for some time
and previous work has already demonstrated the feasibility of detecting bolides using an algorithmic approach
\citep{rumpf-glm}. This work utilized classic filter design where human-derived heuristics are considered to create a
set of detection filters tuned to what a human would expect for a bolide event in GLM data.  The clustered L2 group data
was passed through each filter sequentially and the product of the individual filter scores formed the cumulative
candidate score. If the cumulative candidate score was above the threshold value a detection was recorded. Much effort
was placed on the design of these filters, or `features' in Machine Learning parlance, with each feature identifying a
distinct aspect of the data. Although an excellent first step, the classic filter design had too much bias and too
little variance in the classification logic, resulting in too many false positives.  Being composed of a collection of
hard boundary yes/no decisions, the classic filter pipeline was unable to represent the optimal decision boundary in the
feature space.  In other words, there was too little variance in the model to properly distinguish bolides from false
positives and so the model under-fitted.

\begin{figure}
	\centering
		\includegraphics[scale=.5]{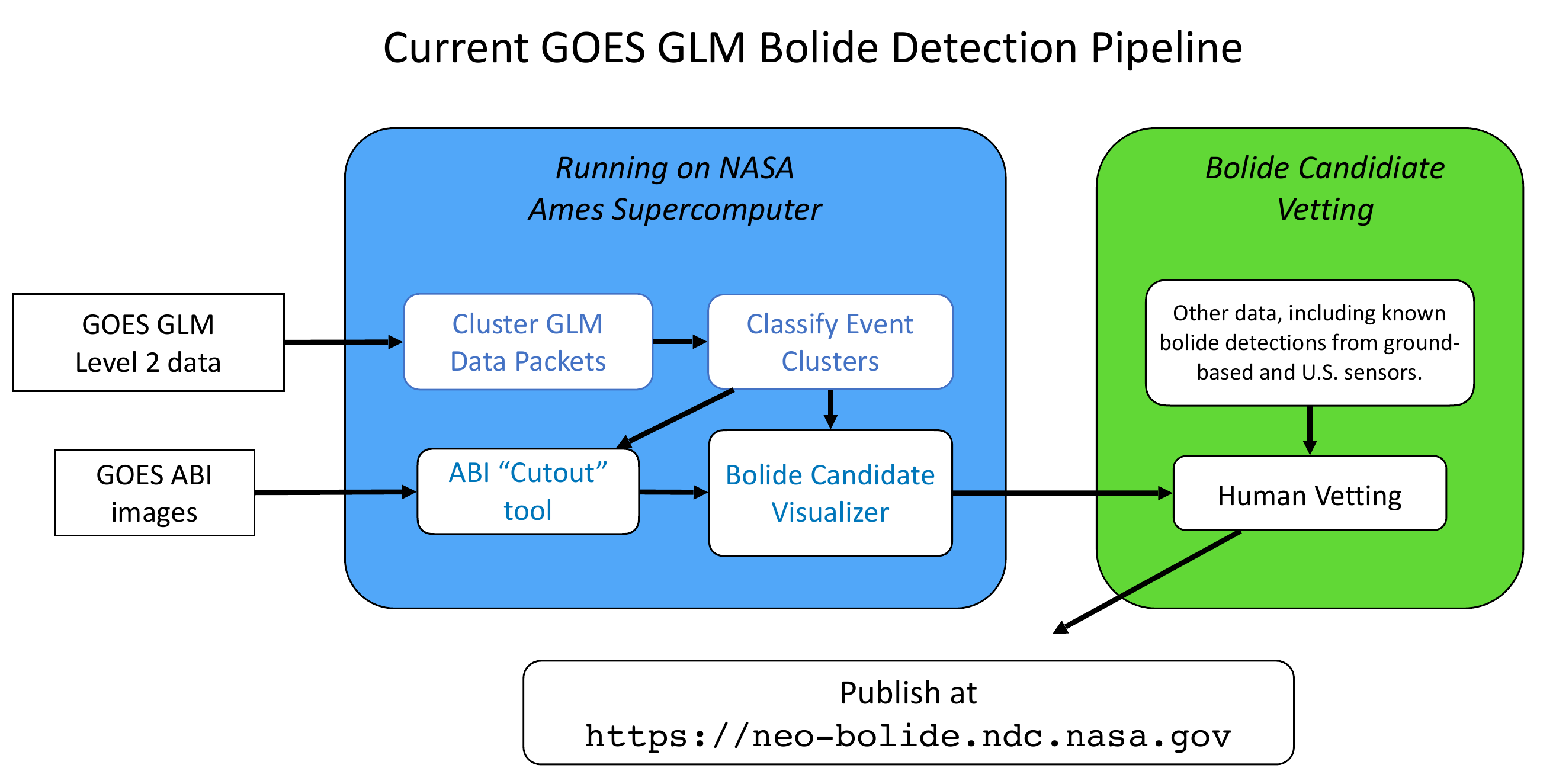}
        \caption{Flowchart of current GLM bolide pipeline.}
	\label{f:currentFlowchart}
\end{figure}

The algorithms developed in \cite{rumpf-glm} were used as a basis for a first generation GLM bolide detection pipeline.
Figure \ref{f:currentFlowchart} gives a flow chart of this pipeline. The pipeline runs on each GLM 20-second Level 2
netCDF file.  We have a local mirror of the GOES 16 and GOES 17 GLM L2 data files, receiving the files within 1 minute
of their generation.  The GLM group data is utilized for detection, as was used in the filter prototype. The GLM data groups are first
clustered into potential bolide detections. The clusters are then classified into bolides and not bolides.
The clusters that pass the classifier have a figure generated similar to Figure \ref{f:bolideExample}. 
The detection candidate
information is then passed to the GOES Advanced Baseline Imager (ABI) cutout tool which identifies a section of a GOES
ABI image data in numerous bands to examine evidence if clouds and weather activity in both day and night 
settings is present around the bolide
candidate. A further figure is also generated which shows all GLM event data within a region around the candidate. All
generated figures are then packaged up and collected for manual perusal by the human vetters. Those candidates that pass
vetting are deemed ``discovered'' bolides and posted on our website. 
The pipeline operates on each 20-second netCDF file independently. If a bolide crossed the
20-second boundary it is detected as two separate bolide candidates, but can be combined in post processing.

\subsection{GLM Group Clustering and Classic Filter Classification}\label{s:clustering}
The first step in the pipeline is the clustering of GLM data packets. The L2
group data in each 20-second netCDF file is clustered using a two-step process as presented in Algorithm
\ref{A:clustering}.
\DecMargin{0.5cm}
\begin{algorithm}\label{A:clustering}
\SetAlgoLined
    \caption{GLM Group Clustering}
    \KwInput{Unsorted GLM groups from a single GLM 20-second netCDF data file}
    \KwOutput{GLM groups clustered into potential bolides}
    \justify
    \begin{algorithmic}[1]
        \STATE \textbf{Hierarchical Clustering:} Uses the time, latitude and longitude distance separation between each
        group and SciPy's \citep{2020SciPy-NMeth} \texttt{cluster.hierarchy.linkage} method, using the nearest point algorithm. 
        \texttt{cluster.hierarchy.fcluster} is then used to form the clusters using the criterion that no clusters contain groups
        with a cophenetic distance greater than $\sqrt{2}$. Time is normalized by 0.2 seconds and distance to a
        Haversine distance of 7.0 km.\label{hierCluster}
        \STATE  \textbf{Sequential Clustering:} For each cluster found in step \ref{hierCluster}, the groups are ordered
        in time then sub-clustered into groups that are
        sequential in lat/lon and time. I.e., sub-clustered into chains of groups. The total time span between
        the first and last group cannot be greater than 0.2 seconds and the greatest Haversine distance not greater
        than 7.0 km.
  \end{algorithmic}
\end{algorithm}

The second step is intended to better identify groups that form chains which would signify a bolide event travelling in space
and time, instead of a lightning event which is more stochastic. This latter point is
the principal reason the GLM \emph{flashes} are not used. The flashes are tuned to cluster groups into lightning-like clusters. 
The 0.2 second and 7.0 km limits can result in some bolides not being completely recovered, but for the detection step
this is acceptable.

Once the clusters of GLM groups are identified, each cluster
is then run through the six filters discussed in \citep{rumpf-glm}. Each cluster that passes the filters is declared
a ``detection,'' recorded and a diagnostic figure is generated.

\subsection{Manual Bolide Candidate Vetting}\label{SS:VettingProcedure}

Until a very high precision detector is deployed, a human-in-the-loop vetting step will be necessary. The initial
filter-based detection discussed above resulted in a considerable number of false positives.  Typically, more than 50
detections were made per satellite every day, sometimes much more than that, especially during high glint activity. 
A more advanced detector, as discussed
below in Section \ref{s:trainedForestClassifier}, has reduced the false positives considerably but a final vetting step
is still necessary. In time we expect the manual vetting steps to be codified in an automated procedure and no
longer need a human-in-the-loop. The vetting begins with a diagnostic figure being generated for each detection. 
A web interface was developed to allow the human vetter to quickly peruse all the detections and
pick out the ones that strike the reviewer as a potential bolide and then pursue further validation investigations on
the smaller triaged set.

The first step in the vetting process consists of a visual inspection of candidate bolides. These figures are similar to
shown in Figure \ref{f:bolideExample}.
This cursory
inspection looks for events that have a well-structured light curve, a linear ground track and a duration of less than a
second\footnote{Very large bolides can have light curves that last well over a second and so if the event is particularly
bright we would not exclude a longer duration.}. 
The determination as to if the bolide candidate should pass triage to the next stage of vetting is based on several key
aspects of the figure as presented in Algorithm \ref{P:vetting}.
\begin{algorithm}\label{P:vetting}
\SetAlgoLined
    \caption{Bolide candidate triage considerations}
    \justify
    \begin{algorithmic}[1]
    \STATE We seek energy light curves that have a well-structured and characteristic shape of a bolide. We desire
        candidates that have: A gradual and even increase in flux intensity, peak in the latter half of the time series
        and then have a more sudden drop at the end.
    \STATE Many bolide candidates with steep trajectories are not detected on multiple GLM pixels, so ground track movement is 
        not always a good metric. But when the track does cover multiple pixels, we assess whether it is linear.
    \STATE Lightning and terrestrial sun glint can sometimes be easily discarded if the data groups cluster more
        stochastically in time and geographic location, instead of gradual and linear progression.
    \STATE Some derived quantities are also presented and considered by the vetter including: Ground track distance, approximate ground
        speed and total duration. These should all have realistic values for a bolide.
    \STATE The classifier detection confidence is also presented, when available. In the three examples in Figures \ref{f:lightningExample}, 
        \ref{f:glintExample} and \ref{f:bolideExample} we see the legitimate bolide has the highest score. It is best
        that the human vetter does not utilize this score so that the classifier does not bias the human. It is nonetheless useful to
        display the score to aid diagnosis of errors in the classifier.
    \end{algorithmic}
\end{algorithm}
For the candidates that pass the initial triage, the next step in the vetting process is to attempt to correlate candidate bolide
events with other detection sources such as the JPL fireball reports (https://cneos.jpl.nasa.gov/fireballs/) or other
ground based observations when a candidate event occurs over land. Ground based observations includes eyewitnesses and
all-sky camera network recordings and are discussed above in the Introduction.
Most candidate bolide events occur over the oceans where ground-based detections are not possible. 

\begin{figure}
	\centering
		\includegraphics[scale=.2]{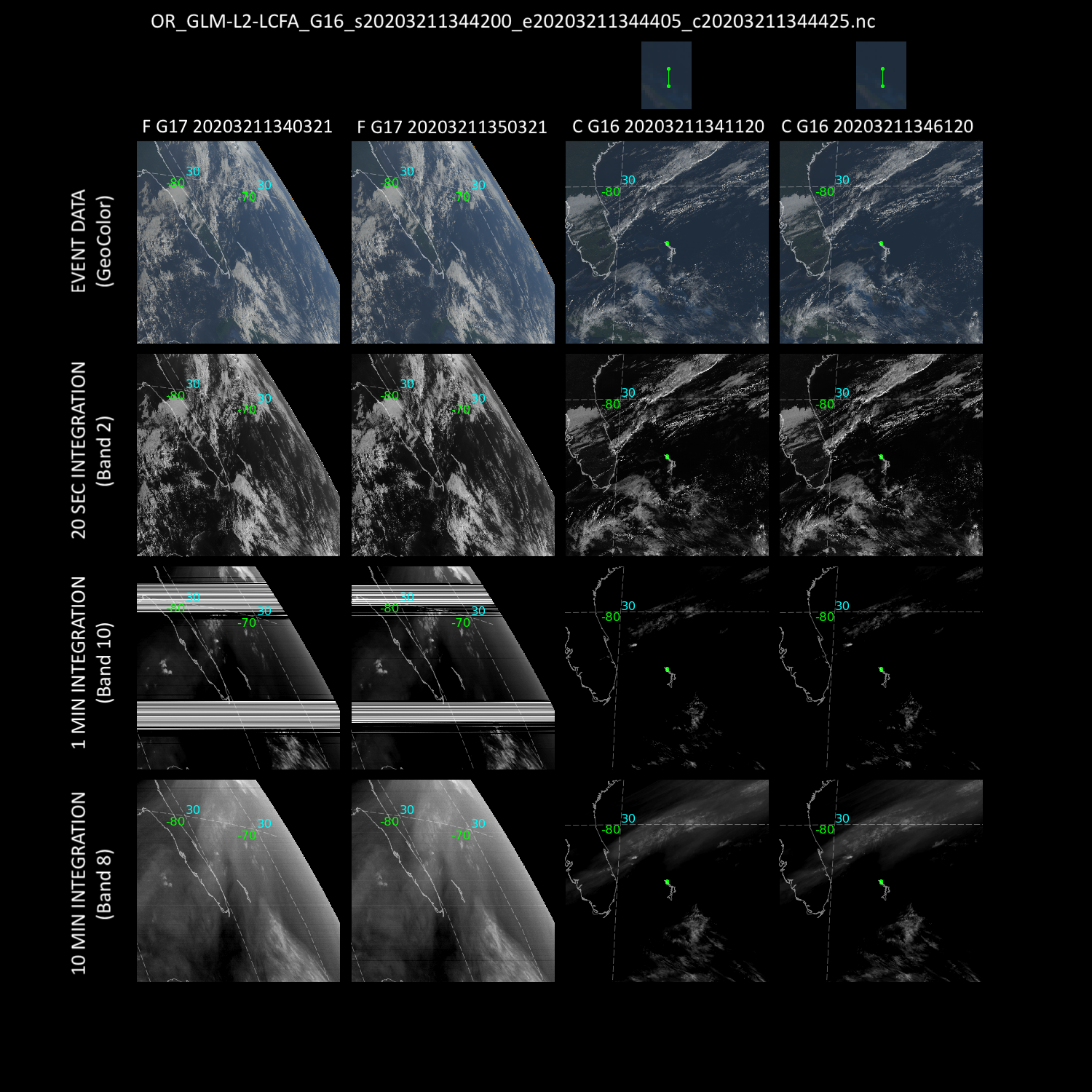}
        \caption{GOES ABI cutout with GLM event data superimposed. The detected bolide is the green dot and corresponds to the bolide in Figure
        \ref{f:bolideExample}. The event was detected in GOES 16 and there is no lightning activity in either satellite
        data, hence, a single green dot in the right two column images. This candidate is therefore highly suspected to be a legitimate bolide.}
	\label{f:bolideCutoutExample}
\end{figure}

The largest source of false positives is lightning activity, which can in many cases look like a bolide.  To eliminate
these sources of false positives, the vetting process relies heavily on overlaying the GLM lightning data on GOES ABI
images.  If a candidate event is isolated from storm activity in time and geographic location then the confidence of it
being a real bolide increases.  Initially, the GLM ABI fusion analysis was performed using a tool developed by Colorado
State University called the Satellite Loop Interactive Data Explorer in Real-time (SLIDER) \citep{MickeSLIDER}.  As we gained
experience examining candidate events, it became clear that critical additional information could be found in the GOES
ABI data and so we developed our own custom tool.  The new tool generates a set of static images that overlay GLM and ABI data to
directly present the most desirable information and to help streamline the vetting process.  During the development of
these static images we investigated different GLM integration times and different ABI bands for cloud assessment.
Figure \ref{f:bolideCutoutExample} shows an example static image that displays 16 separate images in a four by four
matrix. This example corresponds to the bolide detection in Figure \ref{f:bolideExample}.  The geographic extent of each
image is roughly four degrees latitude by four degrees longitude centered about the GLM event location. The actual
geographic extent varies due to the satellite's perspective on each region over the Earth. The first two columns show
ABI data scans for GOES 17 before and after the candidate bolide event time. The second two columns show ABI scans before and
after the candidate event time for GOES 16. The ABI scans are spaced between 5 and 15
minutes apart, depending on which ABI scan mode data is available.
The top row ABI data is the day/night Geo color composite, which composites
bands 1,2,3,7,11,13 (7,11,13 are only shown at night). The second through fourth rows show individual ABI bands.  Band
$2$ (red band) in the second row is the highest spatial resolution. 
Bands $10$ and $8$ in the third and fourth rows respectively, are the
upper- and lower-level water vapor bands and provide a different view from the other bands and have emerged as useful
for showing water vapor signatures.  Stereo data is only shown if the event location is in the stereo region, otherwise
only two columns for a single satellite are shown.   

\begin{figure}
	\centering
		\includegraphics[scale=.2]{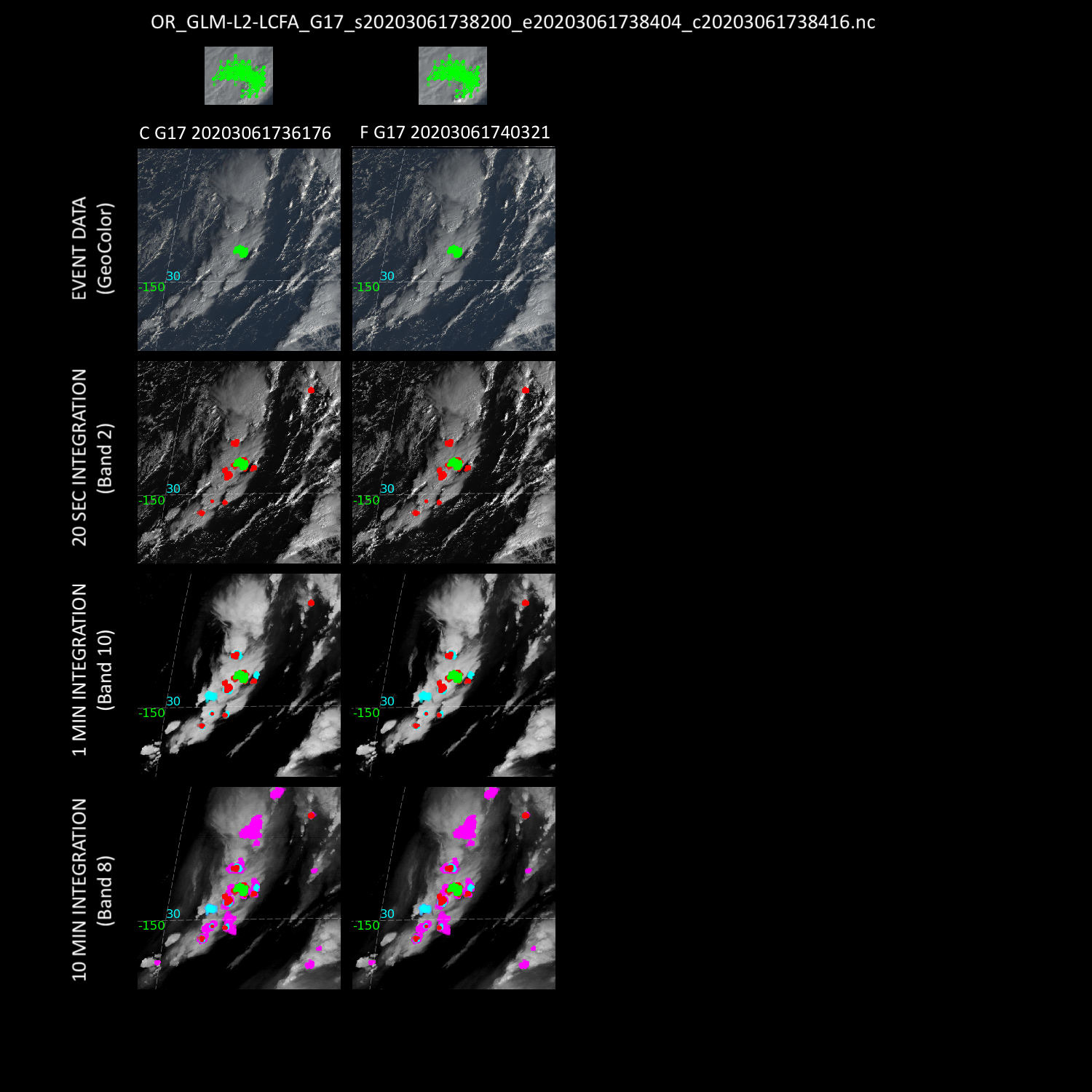}
        \caption{GOES ABI cutout with GLM event data superimposed and corresponds to the lightning in Figure
        \ref{f:lightningExample}. The bolide candidate is the green dot and neighboring GLM lightning activity is in
        red, cyan and magenta. There is no GOES 16 data for this event so only the two left columns are present. We
        see clear cloud and lightning phenomena around the bolide candidate, hence this is highly suspected to be a false
        positive.}
	\label{f:lightningCutoutExample}
\end{figure}

The different rows in the image matrix show different GLM integration times overlaid with the different ABI bands just
explained above. The first row shows only the GLM data associated with the candidate bolide event in green overlaid with
the ABI Geo color composite.  The second row additionally shows the GLM data for the full 20 second GLM data file in
red.  The third row increases the GLM integration time to one minute shown in cyan, so 3 GLM data files with the middle
being the same file as in the previous row. The fourth row shows GLM data for a ten-minute integration time in
magenta, so 30 20-second GLM data files.  This ten-minute integration time is identical to SLIDER’s GLM product. If no
storm activity appears about the bolide detection then the bolide confidence increases.  We can see in Figure
\ref{f:bolideCutoutExample} that the bolide event, which was only detected in GOES 16, is centered in each picture and
there is no other GLM activity or heavy cloud activity within the region of the bolide candidate.  This is clear in the
figure because the GLM data is represented as a single green dot in the right two columns (GOES 16 data) and there is no
other colored data in the figures (no red, cyan or magenta dots). There is also little cloud activity. We thus conclude
this event is highly likely to be a legitimate bolide. A contrasting example is in Figure \ref{f:lightningCutoutExample}
which corresponds to the lightning event in Figure \ref{f:lightningExample}. Here we see a considerable amount of cloud
and lighting activity. There is a large amount of colored GLM data, indicative of lightning within a storm
cloud.  The bolide nature of this event is highly suspect.

Another powerful tool used in both detection and vetting is what we refer to as ``stereo renavigation.'' If there is an
event in the stereo region detected by both GLM-16 and GLM-17 each ground track can be re-navigated by raising the
altitude until ground track parallax errors are minimized. The standard GLM ground segment pipeline \citep{GLM-ATBD}
cannot determine altitude so the GLM Level 2 data product assumes a lightning altitude that is approximately 10 km above
the earth’s surface \citep{10.1117/12.2242141}.  Bolides are luminous at much higher altitudes (usually well above 30km). If
a stereo event intersects at a high renavigated altitude then the confidence is high that the event is a bolide and not
lightning.

Once the vetting process is completed the candidate events are entered into the GLM database using a web-based
interface.  The GLM bolide event date, time, and geographic location are recorded and the corresponding netCDF file
containing the Level 2 event data is uploaded. The public-facing website provides many useful figures and statistics for
each bolide, including a ground track plot, event location on the globe, the energy deposition light curve of the
disintegration and other biographical information for the event. Other comments and observations ascertained by the
human vetter are also recorded, including a human-assessed bolide confidence score.  All vetted GLM bolide events can be
accessed at the NASA Ames GLM Bolide website, https://neo-bolide.ndc.nasa.gov/, which is updated
frequently.

\subsection{Classic Filter Pipeline Performance}\label{s:classifFilterPipeline}
The classic filter detector was a good first step in our pipeline development. It allowed us to find potential bolides,
but with a significant amount of manual vetting required to validate them. The classic filter pipeline would generally
find 100-200 detections per day between GOES 16 and 17. Vetting was laborious but the diagnostic figures discussed above
sped up the process.

Common metrics when assessing the performance of a machine learning classifier are ``precision'' and ``recall''.
Precision means the fraction of detections which are true positives. Recall means the fraction of true positives which
are detected. Since we are running on real data sets, there are no independent
detection methods to verify these bolides, except for a small number of cases,
we therefore have errors in our precision and recall statistics. Recall in particular
is problematic because there are certainly many real bolides we miss. The pipeline first ran periodically in early 2019,
and daily beginning on August 12th, 2019.
Table \ref{T:filterPipelinePerformance} summarizes the bolide detection performance for the filter pipeline between
Sept. 20th, 2019 through November 5th, 2020. Precision is clearly very poor and sifting through the 64,000 total detections was
very laborious.
\begin{table}
    \begin{center}
        \begin{tabular}{| l | c | c |}
            \hline
            \textbf{Filter Pipeline}& G16       & G17       \\ \hline
            Total Detections        & 28,385    & 36,397     \\ \hline
            Total Website Bolides   & 693       & 680       \\ \hline
            Found Website Bolides   & 466       & 466       \\ \hline
            Precision               & 1.6\%     & 1.3\%     \\ \hline
            Recall                  & 67.2\%    & 68.5\%    \\ \hline
        \end{tabular}
        \caption{Summary bolide detection performance for the classic filter pipeline between 
        Sept. 20th, 2019 and November 5th, 2020}
        \label{T:filterPipelinePerformance}
    \end{center}
\end{table}

The recall is merely the fraction of events we have recorded and posted on our website which were found by the pipeline.
The pipeline did not detect all bolides posted on the website mainly for two reasons: firstly, some of the bolides are
found using other discovery methods, such as those listed in the Introduction, and secondly, many bolides are in the
stereo region where GOES 16 and 17 overlap their fields of view. If the bolide was only detected in one or the other in
the stereo region then the bolide is registered as a false negative for the other satellite.  Despite the low precision
of the initial bolide detection pipeline, it allowed us to move from having no detections to having a sizeable database
of detections. We could therefore subsequently begin to explore alternative detection techniques such as supervised
machine learning, which requires a labelled training data set.

\subsection{Training Data Set Generation}
After running the filter pipeline for over one year we collected a sufficient number of confirmed bolides to generate a
training data set for use with more advanced machine learning methods. The true negatives are taken from the clusters
found in Section \ref{s:clustering}.  The clustering step will cluster all GLM data groups, resulting in numerous millions
of clusters; far too many to fit in memory on the computer used for training. Although out-of-core training is a
viable option, for this step we concluded using a random sampling of clusters would be sufficient. We randomly
sampled 2\% of all clusters to generate the true negatives. Because we know we have not detected all bolides in the
data, we assume there to be false negatives in the training data set. We must therefore accept a degree of mislabeled
data. On average, random sampling of clusters resulted in about 2500 clusters per day for GOES 16 and 600 per day for
GOES 17. Due to our observations developing the filter pipeline above, we expect no more than tens of
legitimate bolides per day. The precision in the true negatives data set is therefore greater than 99\% and it is
safe to assume the vast majority of random clusters are true negatives. 
The mislabeled data is therefore small and the training is viable for a random forest classifier, which is
relatively insensitive to mislabeled data \citep{MELLOR2015155, RODRIGUEZGALIANO201293}. 
Table \ref{T:TrainingData} gives a summary of the training data set for both satellites.  The much
smaller number of true negative instances in GOES 17 is a characteristic of the data, not our clustering, which uses the
same algorithm between both satellites.\footnote{Using different clustering algorithms tuned for each satellite is a
viable investigation to be pursued.} The true positives are expected to have higher confidence than the true negatives
due to the true negatives never being manually vetted and some are expected to be mislabeled. Furthermore, we expect
higher energy bolides to have higher confidence. We therefore weight the instances as listed in the table. True
positives' weight is scaled by energy and the true negatives' weight is set at 0.5.
\begin{table}
    \begin{center}
        \begin{tabular}{| l | c | c |}
            \hline
            \textbf{Training Data Set}  & \textbf{G16} & \textbf{G17}       \\ \hline
            Total Instances             & 1,089,411   & 245,665    \\ \hline
            True Positives              & 652         & 658       \\ \hline
            True Negatives              & 1,088,759   & 245,007    \\ \hline
            True Positive Weighting     & \multicolumn{2}{c |}{$\min \left(
                        \left( \frac{\textrm{totEnergy}}{\textrm{medianEnergy}}\right)^{2.0}, 20.0\right)$}     \\ \hline
            True Negative Weighting     &  \multicolumn{2}{c |}{0.5} \\ \hline
        \end{tabular}
        \caption{Training Data Set used for the random forest classifier.}
        \label{T:TrainingData}
    \end{center}
\end{table}

A further note on the generation of the training data set is that the detected bolides reported by the filter
pipeline did not always fully capture the full set of GLM groups for that bolide. During the manual vetting in
Subsection \ref{SS:VettingProcedure} we would always re-find the GLM groups associated with each bolide. When
generating the training data set we then used these re-found bolide groups, instead of the sometimes more limited group set
found by the filter pipeline.

\section{Training the Bolide Classifier}\label{s:trainedForestClassifier}

With a good training data set we could begin training a machine learning classifier. This section discusses the various
steps in training the second iteration classifier.

\subsection{Data Conditioning and Features}\label{ss:dataAndFeatures}
With the initial pipeline using a classic filter classifier it was natural to use the filters as an initial set of
features for a machine learning classifier. But after previous training exercises we identified several classes of
false positives and were able to develop three more features to tackle these. The total set of 9 features is listed in
table \ref{t:features}. 
The name in the second column is the name for the filter given by Table $2$ in the original prototype detector 
paper \citep{rumpf-glm}.
\begin{table*}
    \begin{center}
    \begin{tabular}{|l | r | r |}
    \hline
        \textbf{Feature} & \textbf{Reference \citep{rumpf-glm} Name} & \textbf{Normalization} \\ \hline
        Number of GLM groups            & Group Count   & RobustScaler \\ \hline
        Time duration of light curve    & Duration      & RobustScaler \\ \hline
        Linearity of travel in Latitude/Longitude Metric \#1 & Line Fit & RobustScaler \\ \hline
        Linearity of travel in Latitude/Longitude Metric \#2 & Line Distance & RobustScaler \\ \hline
        Energy distribution of light curve  & Energy Balance & StandardScaler \\ \hline
        Polynomial fit to light curve       & Polynomial Fit & RobustScaler \\ \hline
        Integrated light curve energy       &                & RobustScaler \\ \hline
        Smoothness of light curve           &                & RobustScaler \\ \hline
        Distance to glint point             &                & RobustScaler \\ \hline
    \end{tabular}
    \caption{Features used in the random forest classifier. Column 2 gives the name in the original work. 
        Column 3 gives the Scikit-Learn's \texttt{ColumnTransformer} normalization method.} \label{t:features}
    \end{center}
\end{table*}
The details for the first six will not be repeated here but each filter is kept in its raw score state instead of applying the sigmoid
function given in equations $1$ and $2$ in \citep{rumpf-glm}. The sigmoid in the original work is unnecessary 
because the machine learning
classifier will perform its own calibration. Feature normalization was performed via Scikit-Learn's \citep{scikit-learn}
\texttt{ColumnTransformer} method. The three new features are discussed below.

\subsubsection{Feature: Integrated Light Curve Energy}
This feature is simply the sum energy over all groups in the bolide candidate cluster. This feature was added because it
is generally true that if clusters in Section \ref{s:clustering} are found with very small total energy, then the legitimacy of
the bolide is in greater question.

\subsubsection{Feature: Smoothness of Light Curve Metric}
\begin{figure}
	\centering
		\includegraphics[scale=0.6]{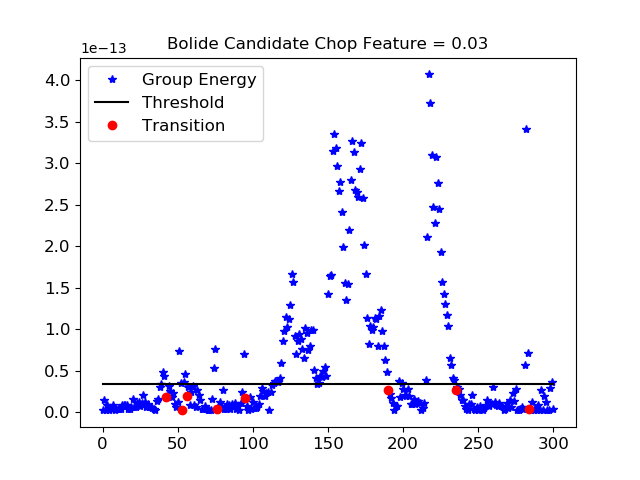}
        \caption{Example calculation of the Smoothness of Light Curve feature for a representative lightning example.
        Note the large number of times the light curve ``transitions'' below a threshold value.}
	\label{f:lightningChopExample}
\end{figure}

\begin{figure}
	\centering
		\includegraphics[scale=0.6]{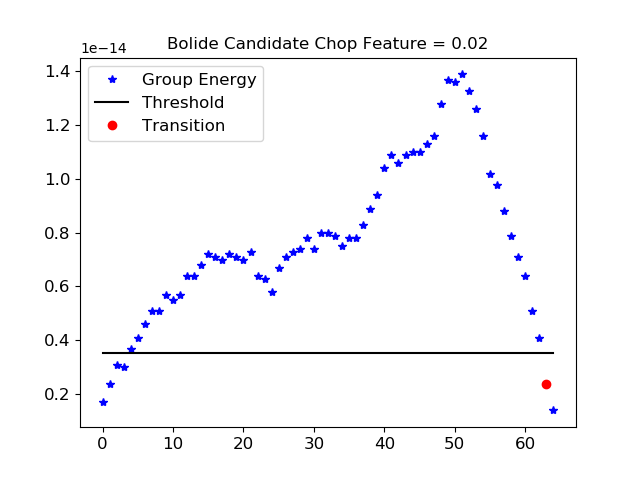}
        \caption{Example calculation of the Smoothness of Light Curve feature for a representative bolide example.
        Note the small number of times the light curve ``transitions'' below a threshold value.}
	\label{f:bolideChopExample}
\end{figure}

Many of the false positives observed are due to a lot of short spikes within the energy profile moving back down to near
the minimum energy. The behavior can be seen in Figure \ref{f:lightningExample} of lightning. This ``choppiness'' in the
light curve is generally indicative of lightning phenomena. Bolides typically have a smoother rise and fall in energy,
and far fewer transitions.  
This feature thus measures the number of occurrences the energy profile first rises above then dips below a
nominal energy value (called a ``transition''). The feature is illustrated in Figures \ref{f:lightningChopExample} 
and \ref{f:bolideChopExample} corresponding to
the lightning and bolide in Figures \ref{f:lightningExample} and \ref{f:bolideExample}, respectively.
The algorithm first calculates a median absolute deviation-based sigma of the light curve, shown as the black line in
the figures.
It then counts the number of times the light curve first rises above 1.1 times the sigma then back below the sigma. 
The feature ignores outliers by only counting a transition if after the energy profile first rises 1.1 times the sigma 
it then drops below the sigma threshold for the subsequent 5 data points. 
It then divides the number of full transitions (red dots in the figures) by the number of groups in
the cluster (blue dots in the figures) to report the feature value.

\subsubsection{Feature: Distance to Glint Point}
The final feature is the distance between the bolide candidate and the glint point. The feature uses the
SpicyPy wrapper to the SPICE toolkit \citep{spicypyAnnex2020}. It first computes the center of the glint point
based on the time and a custom created SPICE kernel for the GOES spacecraft. It then finds the mean latitude and
longitude for each bolide candidate cluster. It finally uses the Haversine formula to compute the great-circle path
length between the glint point and the bolide candidate. 

\subsection{Random Forest Classifier} Simpler machine learning classifiers, such as Logistic Regression
\citep{hastie01statisticallearning} and Support Vector Machines \citep{CC01aSVM}, were tried but results were found to
be inferior to a Random Forest classifier \citep{breimanRandomForests}. Although they did prove to have predictive
power, the former two classifiers lacked the variance inherent in Random Forests and therefore did not have the
versatility to model the feature space as well as Random Forests.  The training instances summarized in Table
\ref{T:TrainingData} and the features in Table \ref{t:features} where used to train a Scikit-learn \citep{scikit-learn}
\texttt{RandomForestClassifier}. 33\% of the data set was reserved for testing. The hyperparameters tuned were
\texttt{n\_estimators}, \texttt{max\_leaf\_nodes} and \texttt{max\_depth}.  Scikit-Learn's \texttt{GridSearchCV} method
was used for hyperparameter optimization with 3-way cross validation and optimization on the F1-Score.  Optimization was
performed on a NAS Pleiades custom node containing a 2-socket Intel Xeon E5-2697 v4 with 18 cores each at 2.30GHz and 
1.5 TB of RAM.
Optimization for each satellite typically took about 1 hour, exclusive of preprocessing time, running $\sim1500$ grid
search points over 50 parallel jobs.  A separate classifier was trained for both GOES 16 and 17. 

AdaBoost boosting \citep{FREUND1997119} was attempted using Scikit-Learn's \texttt{AdaBoostClassifier}
\citep{hastie2009multi} and optimizing again with \texttt{GridSearchCV}, and with adding in the AdaBoost parameters
\texttt{learning\_rate} and \texttt{n\_estimators}.  AdaBoost did not appear to improve performance.

\begin{figure}
	\centering
		\includegraphics[scale=.7]{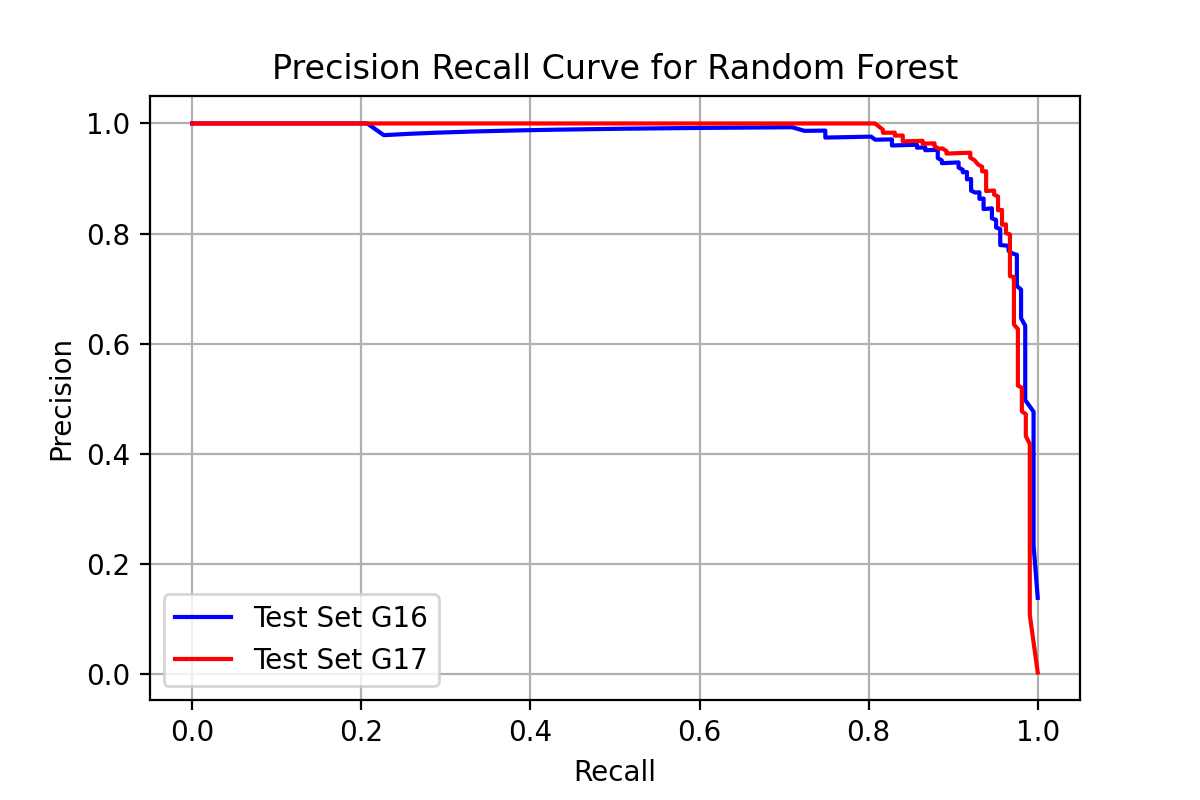}
        \caption{Precision vs. Recall for the Random Forest Classifier on the test set.}
	\label{f:trainingPrecisionRecall}
\end{figure}

Figure \ref{f:trainingPrecisionRecall} presents the Precision vs. Recall curves for GOES 16 and 17 on the reserved test
set. Very good and similar
performance is evident between the two satellites. But note that only 2\% of the clusters were sampled when generating
the true negatives. The training data set is therefore highly unbalanced and one must first scale the precision by a
factor of 50 to obtain the actual precision. Note that the training required over 170 GB of RAM. Using all clusters (50
million for GOES 16 and 12 million for GOES 17) would require out-of-core training or a more memory efficient
algorithm.

One benefit of Random Forests is they graciously provide the gift of feature importance. These are given in Table
\ref{t:featureImportance}. We see that most features have similar importance between the two satellites. The two
discrepancies are ``Number of GLM groups'' and ``Distance to glint point''. This is consistent with the author's manual
investigations of false positives between the two satellites. GOES 17 exhibits more spurious isolated events partly due
to GOES 17's lower detection threshold. A second source is due to a hot pixel on GOES 17 \citep{GLMCDRL038}.
We also see different rates of false positives due to glints between the two
satellites due to GOES 16 being centered over the Americas and GOES 17 centered over the Pacific ocean.
\begin{table*}
    \begin{center}
    \begin{tabular}{|l | r | r |}
    \hline
        \textbf{Feature} & \textbf{GOES 16 Importance} & \textbf{GOES 17 Importance} \\ \hline
        Number of GLM groups            & 0.151 & 0.089 \\ \hline
        Time duration of light curve    & 0.024 & 0.030 \\ \hline
        Linearity of travel in Latitude/Longitude Metric \#1 & 0.192 & 0.211 \\ \hline
        Linearity of travel in Latitude/Longitude Metric \#2 & 0.060 & 0.065 \\ \hline
        Energy distribution of light curve  & 0.011 & 0.010 \\ \hline
        Polynomial fit to light curve       & 0.308 & 0.330 \\ \hline
        Integrated light curve energy       & 0.060 & 0.042 \\ \hline
        Smoothness of light curve           & 0.176 & 0.189 \\ \hline
        Distance to glint point             & 0.019 & 0.035 \\ \hline
    \end{tabular}
    \caption{Random Forest feature importance.} \label{t:featureImportance}
    \end{center}
\end{table*}

\section{Random Forest Pipeline Performance}

\begin{figure}
	\centering
		\includegraphics[scale=.22]{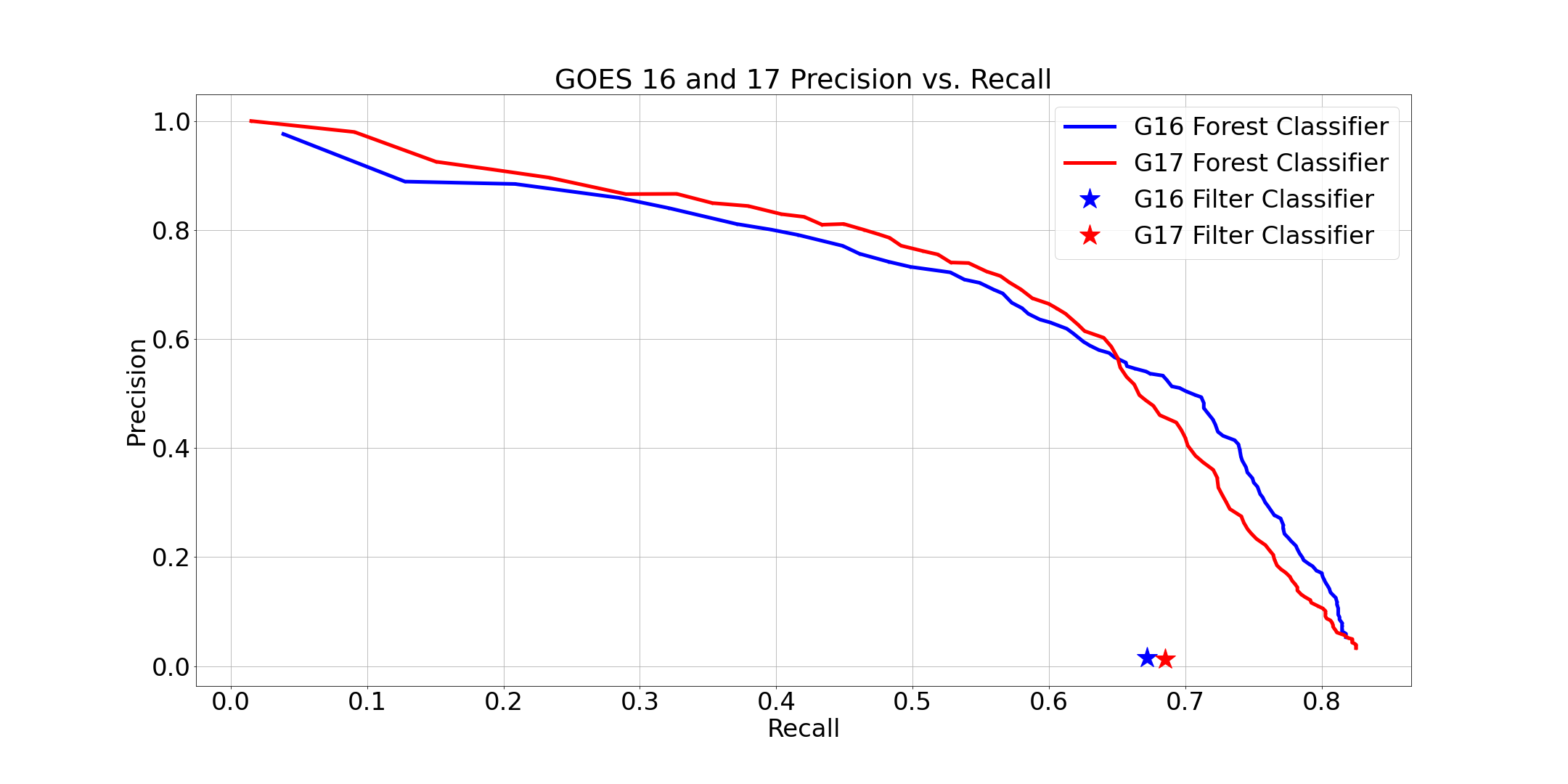}
        \caption{Precision vs. Recall for the full pipeline processing for both satellite pipelines. This is different
        than the classifier training precision vs. recall in Figure \ref{f:trainingPrecisionRecall}. Also note the
        performance of the classic filter pipeline.}
	\label{f:pipelinePrecisionRecall}
\end{figure}

The Random Forest classifier training results in Precision vs. Recall curves as discussed above. However they do not
characterize the real-world performance of the full pipeline. During training, only 2\% of all clusters where used in the
training and test sets. To convert the training precision to real-world precision, we must scale the
precision down by a factor of 50. This imbalanced data set necessary to
train the random forest on a single, albeit very powerful, computer is a major contributor to the discrepancy,
but one must also consider the full pipeline, from
clustering through to human vetting.
To present a more realistic performance metric, Figure \ref{f:pipelinePrecisionRecall} shows the total pipeline 
precision and recall curve as the confidence threshold is varied. 
To illustrate the improvements achieved with the random forest pipeline, we also plot the performance of the classic 
filter pipeline. Note the precision for the classic filter pipeline is near zero.
Obtaining the curves in Figure \ref{f:pipelinePrecisionRecall} required us to reprocess the full data set from 
June 1st, 2019 through
to November, 28th, 2020 but using a very small confidence threshold of 0.01, resulting in a very large number of
``detections.'' Then in post-processing we measured the precision and recall compared to the website data as we
adjusted the confidence threshold. 
Improvement over the classic filter pipeline is evident, but room for improvement is still clearly possible.

\begin{figure}
	\centering
		\includegraphics[scale=.22]{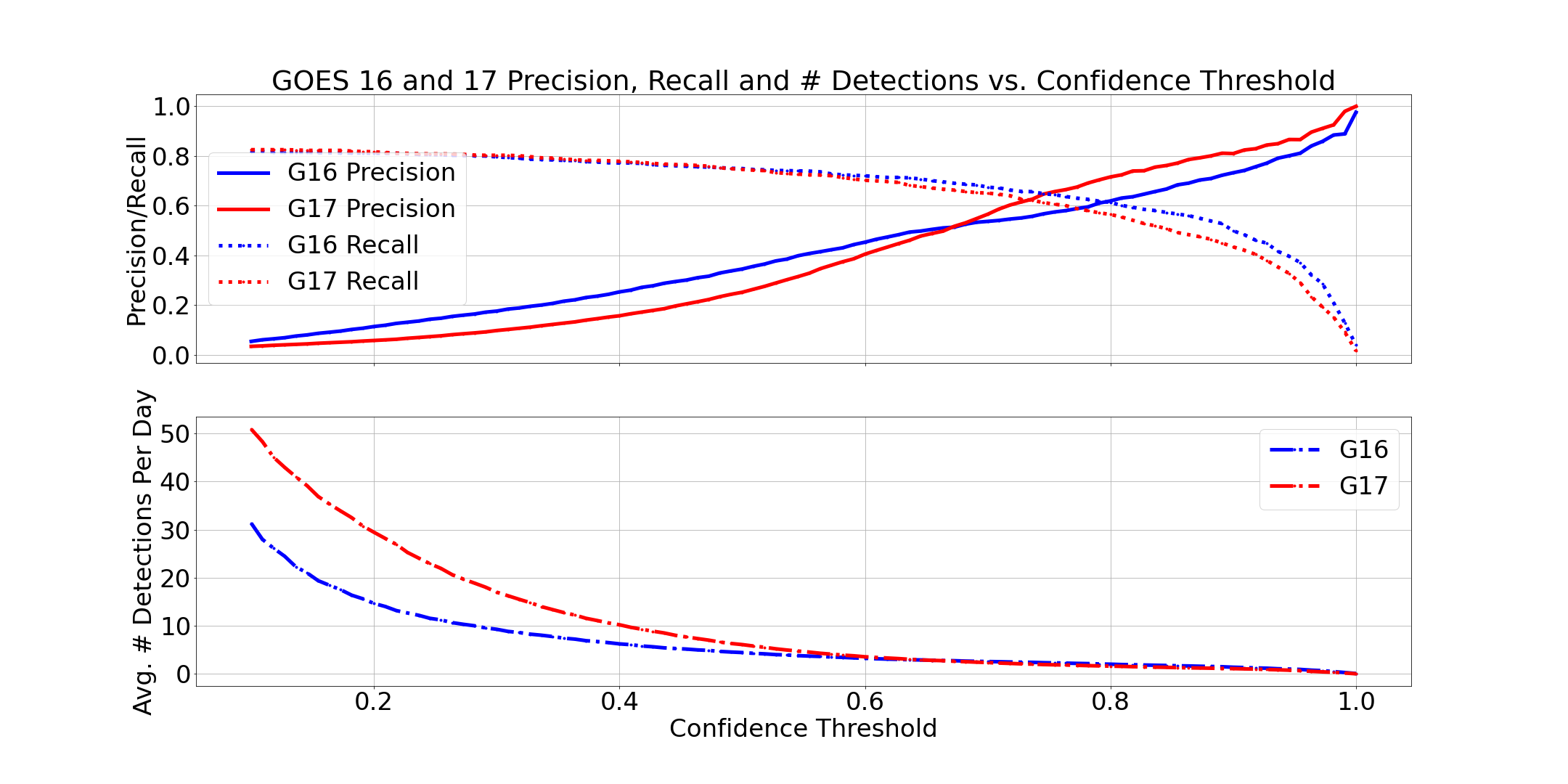}
        \caption{Precision, recall and average number of detections per day versus confidence threshold for both
        satellite pipelines.}
	\label{f:vsConfidenceThreshold}
\end{figure}

One must set the confidence threshold in order to optimize precision and recall. We can see the impact of the
confidence threshold in Figure \ref{f:vsConfidenceThreshold}. Our desire is to have sufficient precision so that the
manual human vetting is not burdensome. A confidence threshold of 0.6 will result in a precision of $\approx 40-45\%$
and a recall of above $70\%$. More importantly, the average number of detections per day would be about 5 -- well within the
requirement to not be a burden to the human vetters.

A summary of the performance of the new pipeline using a confidence threshold of 0.6 is given in 
Table \ref{T:forestPipelinePerformance}. Compared to the filter pipeline in Table \ref{T:filterPipelinePerformance}
we have marked improvement. Precision has improved tremendously, increasing from 1.6\% and 1.3\% to 45.9\% and
41.2\%, respectively. This means, roughly every other pipeline detection is an already known bolide listed on the
website. Recall also improved slightly from 67.1\% and 68.5\% to 74.9\% and 72.4\%, respectively. 
As noted in Section \ref{s:classifFilterPipeline}, measuring the true precision and recall is
problematic due to the involved process to manually vet detections. 
The precision and recall numbers listed in Table \ref{T:forestPipelinePerformance} are therefore lower limits. In order
to determine the true precision and recall we would need to repeat the manual vetting over the entire reprocessed data
set.
\begin{table}
    \begin{center}
        \begin{tabular}{| l | c | c |}
            \hline
            \textbf{Forest Pipeline}& \textbf{G16}  & \textbf{G17}       \\ \hline
            Total Detections        & 1715      & 1880      \\ \hline
            Total Website Bolides   & 1065      & 1070      \\ \hline
            Found Website Bolides   & 788       & 775       \\ \hline
            Precision               & 45.9\%    & 41.2\%    \\ \hline
            Recall                  & 74.0\%    & 72.4\%    \\ \hline
        \end{tabular}
        \caption{Summary bolide detection performance for the random forest pipeline between 
        July. 1st, 2019 and November 28th, 2020}
        \label{T:forestPipelinePerformance}
    \end{center}
\end{table}

\begin{figure}
	\centering
		\includegraphics[scale=.33]{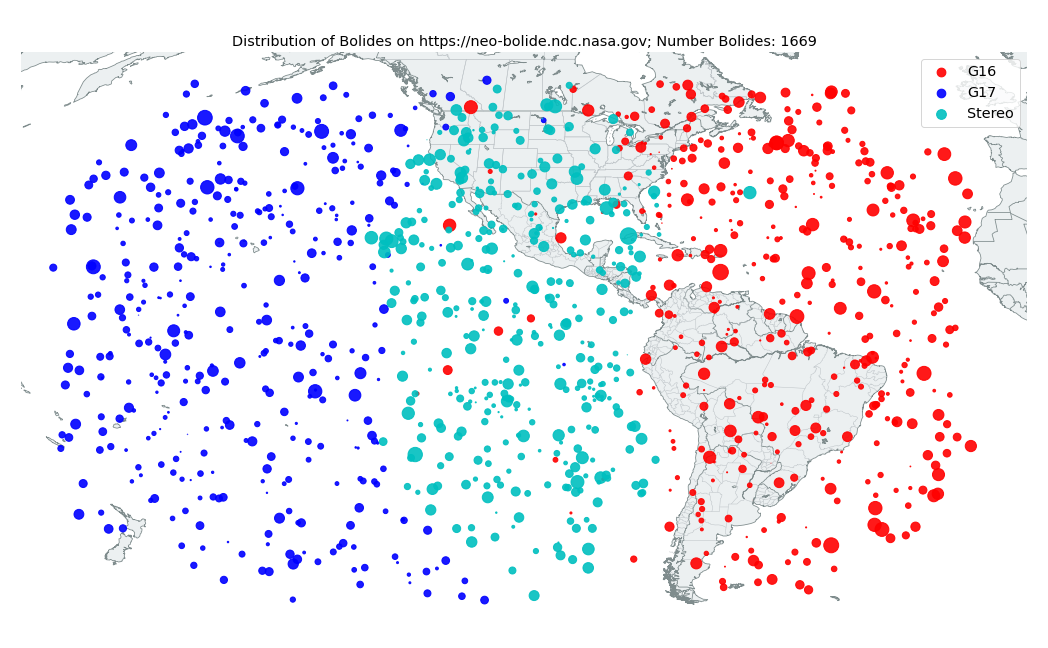}
        \caption{Distribution of 1669 bolides listed on the website as of 2020-11-28. 
        Most are detected via the pipeline but a small number are found via other means.}
	\label{f:websiteBolidesOnGlobe}
\end{figure}

We show the distribution of bolides listed on our website in Figure \ref{f:websiteBolidesOnGlobe}. All bolides shown
are in the GOES GLM data but not all were found in the pipeline. A small number were found by other means but in those
cases the GLM
data was re-examined and if the data was present they were posted to the website, with appropriate comments attached to each
bolide. Bolide total integrated energy is proportional to dot size. The reader is reminded that the plotted energy is that
reported by the GLM ground-segment and therefore not yet calibrated for bolides. Those present in
GOES 16 are in red, those in GOES 17 in blue and those present in both in cyan. A small number in the stereo region have
observable data in only one of the satellites and show up as only blue or red within the cyan dots. The
astute observer will also notice a single cyan dot well east of the North American coast. This bolide was indeed observed
by both instruments when GOES 17 was in its staging location at -89.5 Degrees East before being parked in its final
location at -137.2 Degrees East in November, 2018\footnote{GOES spacecraft latitude and longitude are recorded in the
GLM data files with the keywords \texttt{nominal\_satellite\_subpoint\_lat} and
\texttt{nominal\_satellite\_subpoint\_lon}.}.

\begin{figure}
	\centering
		\includegraphics[scale=.45]{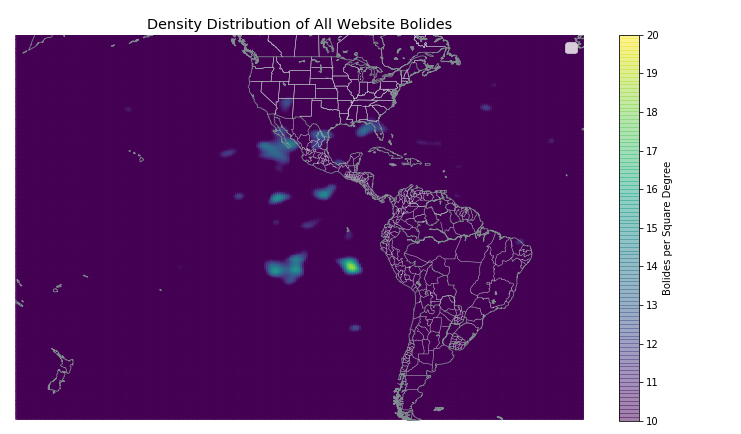}
        \caption{Geographic density distribution of 1669 bolides listed on the website as of 2020-11-28. The colorbar is
        in units of number of bolides per square degree.}
	\label{f:websiteBolidesOnGlobeDensity}
\end{figure}

We can also examine the geographic density distribution of all website bolides in Figure \ref{f:websiteBolidesOnGlobeDensity}. The
distribution is reasonably even however, there is an overabundance within the stereo region. The GOES 16 and 17 pipelines
run independently and there is no difference in the instrument within the stereo region. (Although there
are differences in the detection algorithms between GOES 16 and 17.) We therefore 
suspect the overabundance in the stereo region is due to a
human selection bias: events detected by both instruments within the stereo region
are systematically assigned higher confidence values by the human vetter.
One could argue this is appropriate and if a bolide is detected by two instruments
independently then the event should legitimately have higher confidence. It would be ideal if the overabundance was
evenly distributed over the stereo region, versus the small high dense regions we see indicating a systematic selection bias. 
The variation in bolide impact density over the Earth is an area of active research 
\citep{POKORNY2013682, LEFEUVRE2008291, Evatt2020, robertsonEntryVariation, RUMPF2016209}. Researchers are not yet in
agreement as to how evenly bolides should impact across the globe. However the impact density should only vary with
latitude and not longitude, if averaged over a long time period. With further analysis the data we provide
will aid in these investigations.

\begin{figure}
	\centering
		\includegraphics[scale=.5]{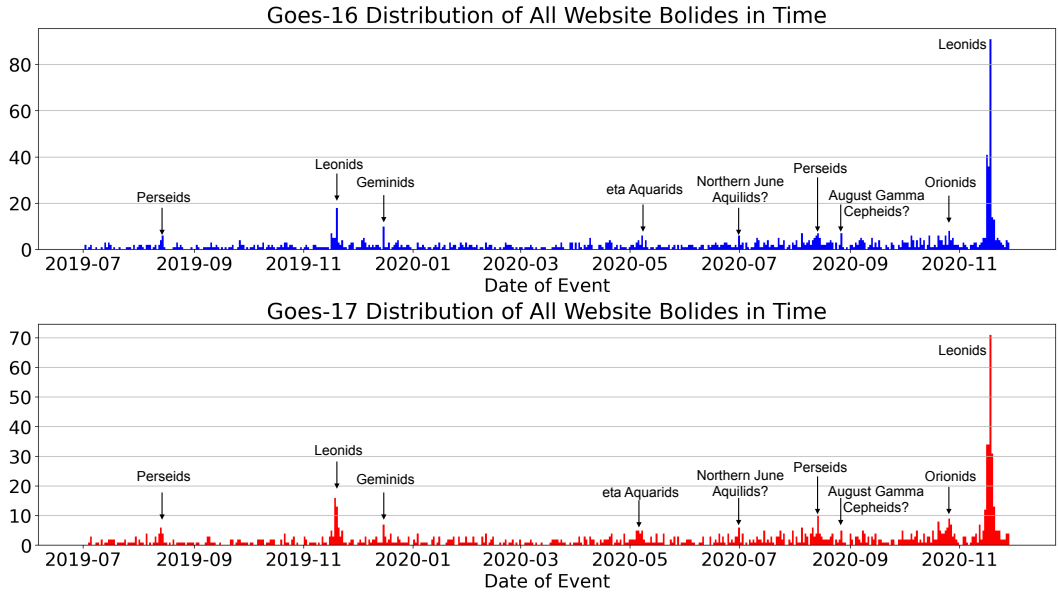}
        \caption{Distribution of 1669 bolides listed on the website as of 2020-11-28. The listed meteor showers line up
        well with the peaks in the histogram.}
	\label{f:websiteBolidesInTime}
\end{figure}

There is also a clear dependence of bolide detection frequency with time, as shown in Figure
\ref{f:websiteBolidesInTime}. Peaks corresponding to known meteor showers \citep{AMS2020Meteors} are highlighted. We
appear to be detecting many of the major showers and perhaps some of the weaker showers. There has been no further
analysis of this figure yet but the alignment with peak shower dates listed by the American Meteor Society strongly suggests we
are indeed detecting them.  The Leonid meteor showers in November, 2019 and 2020 are very clear. It is not yet known why
so many more Leonids were detected in 2020 than 2019. Note that the detection and vetting process has evolved over the
year and a half.  Further investigations are therefore needed to determine if this is selection bias or a true increase
in detectable meteoroids in 2020. 

\begin{figure}
	\centering
		\includegraphics[scale=.45]{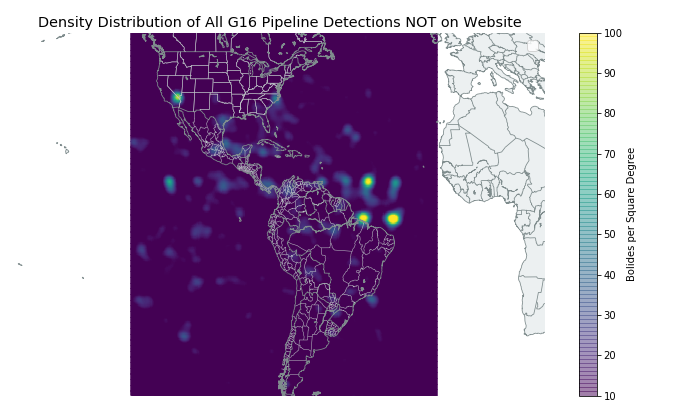}
        \caption{Density heat map of the distribution of all GOES 16 vetting rejections. The dense areas are
        predominantly glint.}
	\label{f:G16PipelineRejections}
\end{figure}

\begin{figure}
	\centering
		\includegraphics[scale=.45]{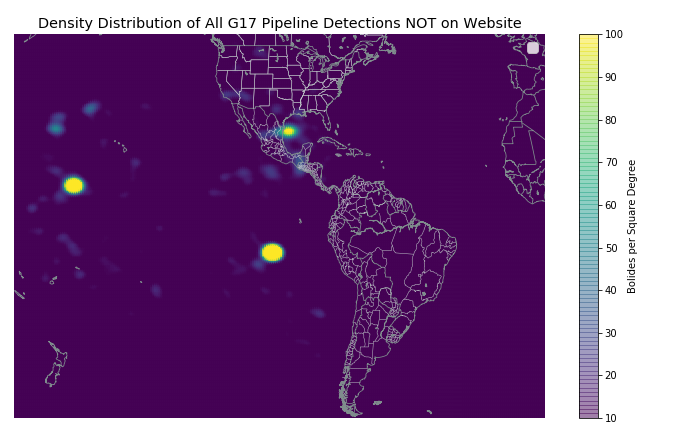}
        \caption{Density heat map of the distribution of all GOES 17 vetting rejections. The dense areas are due to a
        hot pixel on the GOES 17 GLM CCD.}
	\label{f:G17PipelineRejections}
\end{figure}

As a final analysis we can examine the distribution of all pipeline detections \textit{rejected} by the human vetters
over the globe. We would expect the distribution of legitimate bolides to be approximately
even and
the distribution of vetted bolides in Figure \ref{f:websiteBolidesOnGlobeDensity} is indeed close to evenly distributed. 
There
does however exist some irregularities when examining all detections rejected by the human vetter. 
Figures \ref{f:G16PipelineRejections} and \ref{f:G17PipelineRejections} shows a density heat map of all pipeline events
rejected by the human vetters. Note the different colorbar scale versus Figure \ref{f:websiteBolidesOnGlobeDensity}.
We see clear irregularities. For GOES 16 we see a clear overabundance of detections in latitudinal bands near the
equator. There is also a bright spot in southern Nevada and western California, where a large number of solar power
plants are located. These are probably due to glints. For GOES 17
the overabundance is concentrated in two points in the Pacific Ocean. These are due to a documented hot pixel cluster \citep{GLMCDRL038}
on the GOES 17 GLM CCD (GOES 17 undergoes a 180 degree rotation twice a year, hence the two spots).
There is also a slight overabundance over the eastern Gulf of Mexico, which is probably due to
glints.

\section{Future Improvements}\label{s:futureImprovements}

The pipeline has seen tremendous improvements over its short history, nonetheless, there are more improvements to be
made. Our ultimate goal is a fully automated pipeline where bolides are detected, light curves generated and published
to the website with no human intervention. In principle, detections could be published within a minute of data downlink
from GOES. Before we achieve this goal we must continue to improve the precision of the detector. Here lies the
principle desired improvement to the pipeline. There are many avenues to increase precision. The naive method would be
to lower recall until the precision threshold is reached. Figure \ref{f:vsConfidenceThreshold} indicates this strategy
is not promising. We see that the recall would have to drop to near zero in order to have a 95\% precision. We therefore
need to improve the algorithm to reach high precision.

We use a very limited amount of information from GLM to classify our clusters. We only look at the clustered data
itself, but it would be beneficial to also look at the neighboring GLM data groups in both time and latitude/longitude.
We can often identify false positives if there are many GLM groups nearby the bolide candidate. An algorithm could make
this determination. The GOES ABI cutout in Figure \ref{f:bolideCutoutExample} also contains a wealth of weather
information crucial to determine the legitimacy of bolide candidates and should be included. Moving even further abroad,
we could include any other supporting information where a bolide was detected using different methods, such as those of
numerous ground-based detection programs. However, this latter data is not expected to be timely available, and so might
not be able to be used in an automated pipeline with prompt detection. We could also move to using raw level 0 data for
detection.  We anticipate this not being necessary because the bolides are clearly present in the Level 2 data, albeit
perhaps less complete than in Level 0 since some pixel events are discarded by the ground processing to produce Level 2
data. It may be that some bolide signatures are removed entirely by the ground processing. So, maximizing recall may
require Level 0 data. In any event, we intend to begin using the raw Level 0 data when generating the calibrated light
curves because reprocessing the raw data specifically for bolides will result in more accurate bolide light curves.

Many studies have been undertaken to study solar glint in space-based Earth observatories
\citep{petersonGlint,GLM-SANSIA-GLINTS, kayGlints}. We know that glint is a
significant source of background for us. Incorporating some of these techniques could also result in large increases in
precision. We also see the hot pixel on GOES 17 in Figure \ref{f:G17PipelineRejections}. This should be relatively easy to remove.

The third iteration of the classifier is expected to be a convolutional neural network. As we move away from manually
created heuristics and filters and move to allow the machine learning to learn the features on its own it is advantageous
to move to a neural network. We will be able to incorporate other data streams, such as GOES ABI into the neural
network, perhaps even use both GOES satellites simultaneously within the stereo region.  Reprocessing of the full
historical data set is performed with 100 NAS Pleiades Broadwell nodes (28 core Intel Xeon with 128 GB RAM)
and is achieved in about 2 hours. This massively parallelized reprocessing allows for rapid prototyping.

Right now, the confidence score presented on the website is based on the assessments of the human vetters.  There are
three scores: ``low,'' ``medium'' and `high.'' We fully expect many bolides listed on the website to be false positives.
When a ground-based event detection is available to verify our event we make a note on the website, however the process
is very time consuming. We hope to automate this process. In time, we also desire to perform a statistical analysis of
the detection certainty, perhaps augmented with simulated data, to measure the performance of the pipeline and provide a
numerical probability score for each bolide on the website.

A parallel effort currently under investigation is the ``stereo renavigation'' technique introduced in Section
\ref{SS:VettingProcedure}. This method is proving to be successful and a future paper will detail this work.
The website currently provides basic search functionality to find particular bolide events but we also intend to
provide tools to allow the user to download the bolide database in a computer-readable format so that it can be
automatically loaded into data analysis packages.

\begin{figure}
	\centering
		\includegraphics[scale=.5]{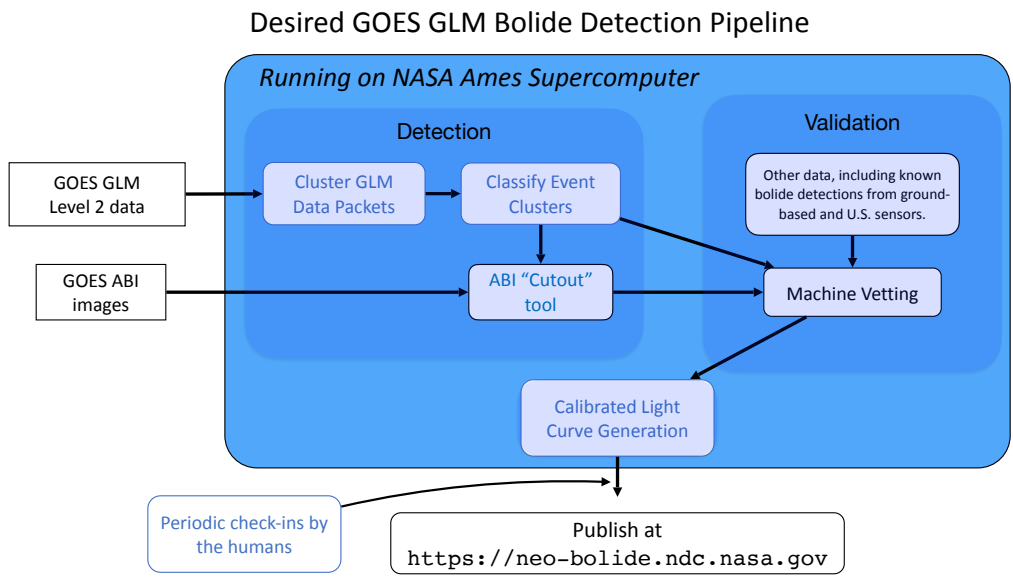}
        \caption{Flowchart of desired GLM bolide pipeline.}
	\label{f:desiredFlowchart}
\end{figure}

Figure \ref{f:desiredFlowchart} gives the desired GLM bolide pipeline. The key difference compared to the current
pipeline in Figure \ref{f:currentFlowchart} is that here the full pipeline is operated automatically on the NAS Pleiades
Supercomputer. This means the "Validation" step is automated. In time we intend to codify the methods used by the human
vetters and perform the analysis, perhaps augmented with a machine learning algorithm, to validate the bolide detections
automatically. Once the detections are validated, we further intend to automatically generate calibrated light curves
and publish them on the website.

\section{Conclusions}

We presented a new bolide detection pipeline for use with the GOES GLM instruments. The work involved taking a prototype
detection algorithm, improving for speed and efficiency and integrating it into an automated pipeline infrastructure,
utilizing a machine learning-based methodology.  The new pipeline runs daily on the NAS Pleiades Supercomputer. The
bolides reported on the website are a product of an evolving pipeline and therefore the distribution of bolides on the
website are subject to a changing selection bias. We have the capability to fully reprocess the historical data set to
generate a consistent bolide catalogue and a full reprocessing can be achieved within a couple hours, utilizing the
massively parallel nature of the supercomputer. These consistently processed data sets may be more useful for
statistical analysis and bolide population studies. We do not yet have the capability to publish these reprocessed lists
on the website, which is reserved for fully human vetted candidates, but hope to make these consistently processed
catalogues available in the future.

The current pipeline bolide detection performance results in a detection precision of 45.9\% and 41.2\% for GOES 16 and
17, respectively. This precision is sufficient for use with human vetters to obtain a reliable
bolide list for publication on our website. However for an automated pipeline with no human-in-the-loop inspection,
higher precision is required and will be pursued as discussed in the Future Improvements section above.

The primary use of our data set is to generate a set of calibrated bolide entry light curves to tune spacecraft reentry
modelling software for use with bolides. But other uses are also apparent.  We can clearly identify common meteor
showers in our data and a statistical analysis of this distribution could aid in understanding these meteor showers.
Once we have a well performing method to analyze the light curves using the reentry modelling software, we could begin to
study the bulk compositions of meteoroids and study the difference in bulk compositions between various showers. 
Within the stereo overlap region viewed by both GOES 16 and 17, it is possible to reconstruct the
trajectory of the incident bolides. Such reconstructions can aid in both investigations of the origins of the bolides and
also increase the prospects of potential meteorite recovery.  As we identify and remove any biases in the dataset, our GLM
detections will become useful in measuring the prevalence of asteroid impacts as a function of latitude. This data can
be used to validate asteroid population models \citep{GRANVIK2018181}, which in turn have much larger implications such
as with regard to our understanding of the evolution of our Solar System.

\section{Acknowledgements}

Work carried out for this project is supported by NASA's Planetary Defense Coordination Office (PDCO). JS and RM are
supported through NASA Cooperative Agreement 80NSSC19M0089. NASA Ames' Code TN division, which operates the
NASA Advanced Supercomputing facility, graciously provided office space and supercomputing credits for our use. We
especially thank TN Deputy Director Donovan Mathias for his support. We also greatly appreciate the help provided by
Lockheed Martin Space System's GLM data algorithms team including Clem Tillier and Samantha Edgington. Additionally, we would
like to thank some very useful discussions about bolide emission spectra with Eric Stern and Darrel Robertson. Finally,
we would like to thank the valuable comments by the two peer reviewers.


\printcredits


\bibliographystyle{model1-num-names}

\bibliography{glm_paper_refs}



\end{document}